\documentclass[aps]{revtex4}

\newcommand{\stac}[2]{\stackrel{\scriptscriptstyle {#1}}{#2}}

\begin{document}

\title{Can we live on a D-brane? ~~-- Effective theory 
on a self-gravitating D-brane --}

\author{Tetsuya Shiromizu$^{(1,3)}, $Kazuya Koyama$^{(2)}$, Sumitada Onda$^{(1)}$, and Takashi Torii$^{(3)}$}

\affiliation{$^{(1)}$Department of Physics, Tokyo Institute of Technology, Tokyo 152-8551, Japan}

\affiliation{$^{(2)}$Department of Physics, The University of Tokyo, Tokyo 113-0033, Japan}

\affiliation{$^{(3)}$Advanced Research Institute for Science and Engineering,
Waseda University, Tokyo 169-8555, Japan}

\begin{abstract}
We consider a D-brane coupled with gravity in type IIB supergravity on 
$S^5$ and derive the effective theory on the D-brane in 
two different ways, that is, holographic and geometrical projection methods. 
We find that the effective equations on the brane obtained by 
these methods coincide. The theory on the D-brane described by 
the Born-Infeld action is not like Einstein-Maxwell theory in the 
lower order of the gradient expansion, 
i.e., the Maxwell field does not appear in the theory. 
Thus the careful analysis and statement for cosmology on 
self-gravitating D-brane should be demanded in realistic models. 

\end{abstract}


\maketitle


\section{Introduction}
The discovery of the D-brane in the string theory has altered the notion of 
extra-dimensions dramatically. The gauge fields are confined 
on the brane and only the gravity can propagate in the whole
higher dimensional spacetime. 
Inspired by this possibility, Randall and Sundrum constructed a 
model where the size of the extra-dimension can be infinite \cite{BW}.
We need no longer a compatification of extra-dimension. Since then, 
the concept of braneworld has been explored intensively in cosmology 
and gravity \cite{BWReview}. 

Although the braneworld is motivated by the D-brane, the 
connection between the D-brane and the braneworld
in Randall-Sundrum model is quite uncertain. 
In the Randall-Sundrum model, gravitational interactions
play an essential role. A self-gravity of the brane is 
essential to localize massless gravitons on the brane. 
As for the confinement of the standard model particle, 
their model completely relies on the idea of the D-brane. 
However, the self-gravity of the D-brane is not clearly 
understood. Although aspects of the D-brane in supergravity are known,
they are derived only by  dual pictures of the probe D-brane. 
Related to this issue, the D-braneworld has been discussed with possible 
assumptions\cite{Dbw}: the brane action is supposed to be Born-Infeld 
type\cite{BI} and the bulk field is only the cosmological 
constant. Adopting the Born-Infeld action as the braneworld action, the 
matter fields are automatically included, although
we have assumed so far that the brane action is the Nambu-Goto plus 
four-dimensional matter action. 
This is an important advantage because
the matter contribution to the braneworld is uniquely determined.  
What we want to do in the current paper is 
deriving an effective theory on the gravitating D-brane
in a more realistic superstring theory. 
   
There are several ways to derive an effective theory on a 
probe D-brane. Recently Sato and Tsuchiya showed that the effective action  
for a probe D-brane can be derived by calculating the classical
on-shell action in type IIB supergravity\cite{ST}. They calculated classical
on shell actions by solving Hamilton-Jacobi equation. Then 
the Born-Infeld action is shown to be a solution. Their analysis opens
up a new possibility to derive an effective theory for a gravitating 
D-brane. A similar situation appears in the context of AdS/CFT 
correspondence. From AdS/CFT correspondence, one can derive the 
generating function for boundary CFT by calculating classical 
action in AdS supergravity. If one introduces a cut-off brane in the AdS 
spacetime, a coupling of gravity to CFT emerges as a consequence of 
the breaking of conformal invariance by a cut-off brane. An interesting 
point is that this effective theory for CFT coupled with gravity 
is nothing but the effective theory for Randall-Sundrum braneworld 
which is a gravitating (cut-off) brane in AdS spacetime.
Thus one may expect that an introduction of a cut-off brane in 
calculations of classical on-shell actions in supergravity  
would provide us an effective theory for a dual quantum field theory 
with the coupling of gravity and/or an effective theory for a 
self-gravitating braneworld. If we adopt this point of view
in the calculation of on-shell actions in type IIB supergravity, 
we might be able to obtain an effective theory for the D-brane coupled 
with gravity and/or a self-gravitating D-braneworld. 

We will derive an effective theory by two ways as in the case for 
a cut-off brane in AdS spacetime and compare each other. 
In the first method we will use the holographic conjecture 
in the braneworld\cite{holo1,Giddings,holo2,holo3,Dholo,holo4}
and solve the Hamilton-Jacobi equation, 
whose solution will play a role of the counter-terms. 
Then adopting an AdS/CFT like correspondence, we can derive
an effective theory for the D-brane with the coupling of the gravity. 
The second one is the geometrical approach  developed in Ref.~\cite{Tess} 
(See also Refs.~\cite{Roy,MW,GE,GE2}), 
we can obtain the gravitational equation on the gravitating brane 
by projecting the five-dimensional variables onto the brane, 
while it is not closed in four dimensions. Indeed, the projected 
bulk Weyl tensor appears as a source term. In the vacuum bulk case 
the bulk Weyl tensor will be negligible in the low energy limit while
it will not be in the case when the non-trivial bulk 
fields exist\cite{holo3,Hime,GE2,Misao}. In the holographic point of view, 
only a part of the bulk Weyl tensor behaves as  conformal field theory(CFT) 
on the boundary\cite{holo3,Hime,GE2}. To identify the CFT part of 
the bulk Weyl tensor, we must solve the bulk matter and 
gravitational fields. After that we can derive an effective theory 
for a gravitating brane. We expect that almost the same result is 
obtained in the holographic approach.  

We will work in type IIB supergravity on $S^5$ 
because the AdS/CFT correspondence 
was originally formulated between super Yang-Mills theory and 
type IIB supergravity aided by D-branes\cite{adSCFT}. For simplicity, 
however, we will turn off several fields in the course of calculation. 
The rest of this paper is composed of two main parts. 
In Sec.~II, we will adopt the holographic method. We first 
describe the strategy and then obtain the solution to the Hamilton-Jacobi 
equation. Finally we see the effective theory 
on the D-brane with the coupling to the gravity. 
In Sec.~III, we will solve the bulk in a long-wave approximation
and try to get the effective equation on the gravitating D-brane in the 
geometrical approach. To make this procedure work well, we will put a 
specific ansatz on a 0th order solution. That is, we need an analytical
background solution in order to solve the next order equations. 
In Sec.~IV, we will give discussions. Therein we will compare the  
results obtained in each methods and present their interpretations.


\section{Holographic approach}

We will derive the effective action for gravitating D-brane using the 
AdS/CFT correspondence. 
See Ref.~\cite{Dholo} for the study of holography on {\it probe} D-branes. 
This section
is organised as follows. 
We begin with the Hamilton-Jacobi equation in the Sec.~\ref{subsecIIA} and
give its  solution 
in the Sec.~\ref{subsecIIB}.  Then we derive the effective theory on the
D-brane  following the braneworld AdS/CFT correspondence in the 
Sec.~\ref{effective}.  We will consider two cases where the 
D-brane is
described by Born-Infeld action and supposed to be done by 
Nambu-Goto action.

\subsection{Type IIB Supergravity on $S^5$ and Hamilton-Jacobi equation}
\label{subsecIIA}

We begin with the action for type IIB supergravity on $S^5$:
%
\begin{eqnarray}
S & = & \frac{1}{2\kappa^2}\int d^5 x {\sqrt {-g}}\biggl\{e^{-2\phi
+\frac{5}{4}\rho}
\biggl[{}^{(5)}\!R+4 (\nabla \phi)^2+\frac{5}{4}(\nabla \rho)^2 
-5\nabla \phi \nabla \rho -\frac{1}{2}|H|^2 \biggr] \nonumber \\
& & ~~-\frac{1}{2}e^{\frac{5}{4}\rho} \Bigl[ (\nabla \chi)^2 
+|\tilde F|^2+|\tilde G|^2\Bigr] +e^{-2\phi +\frac{3}{4}\rho}R_{(S^5)}
\biggr\},
\end{eqnarray}
%
where $H_{MNK}=\frac{1}{2}\partial_{[M} B_{NK]}$, 
$F_{MNK}=\frac{1}{2}\partial_{[M}C_{NK]}$, 
$G_{K_1 K_2 K_3 K_4 K_5}=\frac{1}{4!}\partial_{[K_1}D_{K_2 K_3 K_4 K_5]}$, 
$\tilde F = F+\chi H$ and $\tilde G = G + C \wedge H $. 
$|A_q|^2=\frac{1}{q!}A_{K_1\cdots K_q}A^{K_1\cdots K_q}$.
$M,N=0,1,2,3,4$ and 
hereafter we set $2\kappa^2=1$.  
For example, see Ref.~\cite{ST} for the derivation.

Recently Sato and Tsuchiya derived the Born-Infeld action for a probe
D-brane  as a solution to the Hamilton-Jacobi equation\cite{ST}. Since
the effective  action is obtained via the transition amplitude from the
vacuum to the  boundary state representing the probe D3-brane, it could
be classical  counter-terms. The solutions to Hamilton-Jacobi equation
is raised  in the classical limit of Wheeler-De Witt equation. 

In this paper we will consider the self-gravitating D3-brane, not probe one. 
Our purpose is to get the action for the gravitating D-brane where we 
can discuss the cosmology correctly. For this purpose 
we first write down the full expression of the Hamilton-Jacobi equation 
%
\begin{eqnarray}
& & - \frac{e^{2\phi-\frac{5}{4}\rho}}{({\sqrt {-q}})^2} \biggl[ 
\biggl(\frac{\delta S}{\delta q_{\mu\nu}} \biggr)^2
+\frac{1}{2} \biggl(\frac{\delta S}{\delta \phi} \biggr)^2
+\frac{1}{2}q_{\mu\nu}\frac{\delta S}{\delta q_{\mu\nu}} \frac{\delta S}{\delta \phi}
+\frac{4}{5}\biggl(\frac{\delta S}{\delta \rho} \biggr)^2 
+ \frac{\delta S}{\delta \phi}
\frac{\delta S}{\delta \rho} +\biggl(\frac{\delta S}{\delta B_{\mu\nu}}
-\chi \frac{\delta S}{\delta C_{\mu\nu}}
-6 C_{\alpha\beta}\frac{\delta S}{\delta D_{\mu\nu\alpha\beta}} \biggr)^2\biggr] \nonumber \\
& & ~~-e^{-2\phi+\frac{5}{4}\rho}\biggl[{}^{(4)}R+4D^2\phi -\frac{5}{2}D^2\rho 
-4(D\phi)^2 
-\frac{15}{8}(D\rho)^2+5 D\phi D\rho -\frac{1}{12}H_{\mu\nu\alpha}
H^{\mu\nu\alpha} \biggr]-e^{-2\phi +\frac{3}{4}\rho}R_{(S^5)}
\nonumber \\
& & ~~-e^{\frac{5}{4}\rho} \biggl[-\frac{1}{2}(D\chi)^2-\frac{1}{12}\tilde F_{\mu\nu\alpha}
\tilde F^{\mu\nu\alpha} \biggr]
-\frac{e^{-\frac{5}{4}\rho}}{({\sqrt {-q}})^2}
\biggl[ \frac{1}{2}\biggl(\frac{\delta S}{\delta \chi} \biggr)^2 
+\biggl( \frac{\delta S}{\delta C_{\mu\nu}}\biggr)^2
+12\biggl( \frac{\delta S}{\delta D_{\mu\nu\alpha\beta}}\biggr)^2 \biggr]
=0,
\end{eqnarray}
%
where $q_{\mu\nu}$ and $D_\mu$ are the induced metric on the D3-brane 
and its covariant derivative. $\mu ,\nu=0,1,2,3$. 

In Ref.~\cite{ST}, all fields were supposed to be constant and then 
it was shown that the Born-Infeld action with the Wess-Zumino terms is 
a solution up to full orders of $\alpha'$:
%
\begin{eqnarray}
\stac{(0)}{S}_{BI} =  \alpha \int d^4 x {\sqrt {-q}}e^{-2\phi +\rho} 
+\beta \int d^4 x e^{-\phi}{\sqrt {-{\rm det}(q_{\mu\nu}+B_{\mu\nu})}}
+\gamma \biggl( \int D +\int C \wedge B +\frac{1}{2}\int \chi B \wedge B \biggr),
\end{eqnarray}
%
where $\alpha^2 = 5 R_{(S^5)}$ and $\beta^2 = \gamma^2$. 
In our paper, on the other hand, we will not assume these fields are constant. 
To solve the Hamilton-Jacobi equation, we will employ the gradient 
expansion scheme in proceeding sections.

\subsection{Solution to Hamilton-Jacobi equation}
\label{subsecIIB}

Let us solve the Hamilton-Jacobi equation using the gradient 
expansion scheme. The expansion parameter is $\epsilon= \ell^2/L^2$, 
where $\ell$ and $L$ are the bulk curvature scale and 
the typical gradient scale on the brane, respectively. The solution is 
expanded as 
%
\begin{eqnarray}
S=S_0+S_1+S_2+\cdots.
\end{eqnarray}
%
For example, $S_1$ is expected to contain a linear combination 
of ${}^{(4)}\!R$, $B_{\mu\nu}B^{\mu\nu}$, $(D\phi )^2$ and so on.

\subsubsection{0th order}

In the zeroth order the Hamilton-Jacobi equation becomes 
%
\begin{eqnarray}
& & - \frac{e^{2\phi-\frac{5}{4}\rho}}{({\sqrt {-q}})^2} \biggl[ 
\frac{\delta S_0}{\delta q_{\mu\nu}} \frac{\delta S_0}{\delta q_{\alpha\beta}} q_{\mu\alpha} q_{\nu\beta}
+\frac{1}{2} \biggl(\frac{\delta S_0}{\delta \phi} \biggr)^2
+\frac{1}{2}q_{\mu\nu}\frac{\delta S_0}{\delta q_{\mu\nu}} \frac{\delta S_0}{\delta \phi}
+\frac{4}{5}\biggl(\frac{\delta S_0}{\delta \rho} \biggr)^2 
+ \frac{\delta S_0}{\delta \phi}\frac{\delta S_0}{\delta \rho} \biggr] 
\nonumber  \\
& & ~~-e^{-2\phi +\frac{3}{4}\rho}R_{(S^5)} 
-12\frac{e^{-\frac{5}{4}\rho}}{({\sqrt {-q}})^2}
\biggl( \frac{\delta S_0}{\delta D_{\mu\nu\alpha\beta}}\biggr)^2 =0. 
\label{hj0}
\end{eqnarray}
%
It is easy to see that the solution can be written as 
%
\begin{eqnarray}
S_0 =  \int d^4x {\sqrt {-q}} \biggl[\alpha_0 e^{-2\phi+\rho}+\beta_0 e^{-\phi}
+\frac{\gamma_0}{24}\epsilon^{\mu\nu\alpha\beta}D_{\mu\nu\alpha\beta} \biggr].
\end{eqnarray}
%
Substituting the above into Eq.~(\ref{hj0}), we have an equation for 
$\alpha_0$, $\beta_0$ and $\gamma_0$
%
\begin{eqnarray}
\biggl[ \frac{1}{5}\alpha_0^2 -R_{(S^5)}\biggr]e^{-2\phi+\frac{3}{4}\rho}
-\frac{1}{2}(\beta_0^2-\gamma_0^2)e^{-\frac{5}{4}\rho}=0,
\end{eqnarray}
from which we find
%
\begin{eqnarray}
\alpha_0^2=5R_{(S^5)}~~{\rm and}~~\beta_0^2=\gamma_0^2.
\end{eqnarray}
%

\subsubsection{1st order}

In the first order the Hamilton-Jacobi equation becomes 
%
\begin{eqnarray}
& & -\frac{e^{2\phi-\frac{5}{4}\rho}}{({\sqrt {-q}})^2}\biggl[
2\frac{\delta S_0}{\delta q_{\mu\nu}} \frac{\delta S_1}{\delta q_{\alpha\beta}}
q_{\mu\alpha}q_{\nu\beta}
+\biggl( \frac{\delta S_0}{\delta \phi}+\frac{1}{2}\frac{\delta S_0}{\delta q_{\mu\nu}}
q_{\mu\nu}+\frac{\delta S_0}{\delta \rho}\biggr)
\frac{\delta S_1}{\delta \phi} 
+\frac{1}{2}\frac{\delta S_0}{\delta \phi}\frac{\delta S_1}{\delta q_{\mu\nu}}q_{\mu\nu}
+\biggl(\frac{8}{5}\frac{\delta S_0}{\delta \rho}+\frac{\delta S_0}{\delta \phi} 
 \biggr) \frac{\delta S_1}{\delta \rho} \nonumber \\
& & ~~+\biggl(
\frac{\delta S_1}{\delta B_{\mu\nu}}-\chi \frac{\delta S_1}{\delta C_{\mu\nu}}
-6 C_{\alpha\beta}\frac{\delta S_0}{\delta D_{\mu\nu\alpha\beta}} \biggr)^2
\biggr]
-e^{-2\phi+\frac{5}{4}\rho} \biggl[{}^{(4)}R+4D^2 \phi -\frac{5}{2}D^2 \rho 
-4(D\phi)^2  \nonumber \\
& & ~~-\frac{15}{8}(D\rho)^2+5 D\phi D\rho 
-\frac{1}{12}H_{\mu\nu\alpha}H^{\mu\nu\alpha}\biggr] 
+e^{\frac{5}{4}\rho}\biggl[\frac{1}{2}(D\chi)^2+\frac{1}{12}\tilde F_{\mu\nu\alpha}
\tilde F^{\mu\nu\alpha} \biggr] 
- e^{-\frac{5}{4}\rho} \biggl(\frac{1}{{\sqrt {-q}}}
\frac{\delta S_1}{\delta C_{\mu\nu}} \biggr)^2 =0.
\label{hj1}
\end{eqnarray}
%
For simplicity we set $H_{\mu\nu\alpha}=0$ and $\tilde F_{\mu\nu\alpha}=0$. 
Thus $B_{\mu\nu}$ and $C_{\mu\nu}$ are closed, and then  written 
by the vector potentials. We will also set $C_{\mu\nu}=0$ at the end of 
calculations. Using the solution of $S_0$, Eq.~(\ref{hj1}) becomes 
%
\begin{eqnarray}
& & -\frac{e^{2\phi-\frac{5}{4}\rho}}{{\sqrt {-q}}}\biggl[
\beta_0 e^{-\phi}\biggl( \frac{1}{2}q_{\mu\nu}\frac{\delta S_1}{\delta q_{\mu\nu}}
-\frac{\delta S_1}{\delta \rho}\biggr)
-\frac{2}{5}\alpha_0 e^{-2\phi+\rho} \frac{\delta S_1}{\delta\rho }
+\frac{1}{{\sqrt {-q}}}\biggl(
\frac{\delta S_1}{\delta B_{\mu\nu}}-\chi \frac{\delta S_1}{\delta C_{\mu\nu}}
-6 C_{\alpha\beta}\frac{\delta S_0}{\delta D_{\mu\nu\alpha\beta}} \biggr)^2
\biggr]\nonumber \\
& & ~~-e^{-2\phi+\frac{5}{4}\rho} \biggl[{}^{(4)}R+4D^2 \phi -\frac{5}{2}D^2 \rho 
-4(D\phi)^2  
-\frac{15}{8}(D\rho)^2+5 D\phi D\rho \biggr] 
+\frac{1}{2}e^{\frac{5}{4}\rho}(D\chi)^2 
- e^{-\frac{5}{4}\rho} \biggl(\frac{1}{{\sqrt {-q}}}
\frac{\delta S_1}{\delta C_{\mu\nu}} \biggr)^2 \nonumber \\
& &  ~=0.
\end{eqnarray}
%

Here remember that the AdS/CFT correspondence will hold in the limit of 
%
\begin{eqnarray}
\alpha_0 \to 0,
\end{eqnarray}
%
that is, the AdS and $S^5$ curvature radii are much longer than 
the string length. In this limit we can see that the solution for $S_1$ 
is given by 
%
\begin{eqnarray}
S_1& = & \frac{1}{\beta_0}\int d^4 x {\sqrt {-q}}e^{-3\phi+\frac{5}{2}\rho}
\biggl[ \frac{1}{2}{}^{(4)}\!R+4(D\phi)^2 +\frac{35}{16}(D\rho)^2 
-\frac{25}{4} D\phi D\rho \biggr] 
-\frac{1}{4\beta_0}\int d^4 x {\sqrt {-q}}e^{-\phi+\frac{5}{2}\rho}(D\chi)^2
\nonumber \\
& & +\frac{\gamma_0}{4}\int d^4 x {\sqrt {-q}}e^{-\phi}B_{\mu\nu}B^{\mu\nu} 
+\frac{\gamma_0}{4} \int d^4x {\sqrt {-q}}\epsilon^{\mu\nu\alpha\beta}
\biggl[B_{\mu\nu}C_{\alpha\beta}+ \frac{\chi}{2}B_{\mu\nu}B_{\alpha\beta}
\biggr].
\end{eqnarray}
%
Hereafter we will consider the limit of $\alpha_0=0$ and $R_{(S^5)}=0$.

\subsubsection{2nd order}

Next we consider the second order. The Hamilton-Jacobi equation is 
%
\begin{eqnarray}
& &  \frac{1}{{\sqrt {-q}}} \biggl[ 
\frac{1}{2}q_{\mu\nu}\frac{\delta S_2}{\delta q_{\mu\nu}}
-\frac{\delta S_2}{\delta \rho} +B_{\mu\nu} 
\frac{\delta S_2}{\delta B_{\mu\nu}} \biggr] \nonumber \\
& &~~~ =     -\frac{e^{\phi}}{\beta_0({\sqrt {-q}})^2}
\biggl[ 
\frac{\delta S_1}{\delta q_{\mu\nu}} \frac{\delta S_1}{\delta q_{\alpha\beta}} q_{\mu\alpha} q_{\nu\beta}
+\frac{1}{2} \biggl(\frac{\delta S_1}{\delta \phi} \biggr)^2 
+\frac{1}{2}q_{\mu\nu}\frac{\delta S_1}{\delta q_{\mu\nu}} 
\frac{\delta S_1}{\delta \phi} 
+\frac{4}{5}\biggl(\frac{\delta S_1}{\delta \rho} \biggr)^2 
+ \frac{\delta S_1}{\delta \phi}\frac{\delta S_1}{\delta \rho} \biggr]
-\frac{e^{-\phi}}{2\beta_0}
\biggl( \frac{1}{{\sqrt {-q}}}\frac{\delta S_1}{\delta \chi}\biggr)^2 
\nonumber \\
& & ~~~= -\frac{e^{-5\phi+5\rho}}{4\beta_0^3} 
\biggl({}^{(4)}\!R_{\mu\nu}{}^{(4)}\!R^{\mu\nu}+\frac{1}{2}{}^{(4)}\!R^2\biggr)
+\frac{1}{2\beta_0}e^{-3\phi+\frac{5}{2}\rho}
\biggl[{}^{(4)}\!R^{\mu\nu}(B^2)_{\mu\nu}-\frac{1}{4}{}^{(4)}\!R\;{\rm Tr}(B^2) 
\biggr] \nonumber \\
& & ~~~~~~-\frac{3\beta_0 e^{-\phi}}{8} \biggl[ {\rm Tr}(B^4) 
-\frac{1}{4}\bigl({\rm Tr}(B^2)\bigr)^2 \biggr]+\cdots, \label{anomaly}
\end{eqnarray}
%
where $(B^2)_{\mu\nu}=B_{\mu}^{~\alpha}B_{\alpha\nu}$ and ${\rm Tr}(B^2)
=B_{\mu\nu}B^{\nu\mu}$. 

As seen soon, ${}^{(4)}\!R_{\mu\nu} =O(B^2)$ and ${}^{(4)}\!R = O(B^4)$ 
will be held. Bearing this in mind, $S_2$ can be evaluated as 
%
\begin{eqnarray}
S_2 & = & \frac{1}{20\beta_0^3} \int d^4 x{\sqrt {-q}}e^{-5\phi + 5\rho}
\;{}^{(4)}\!R_{\mu\nu}{}^{(4)}\!R^{\mu\nu} 
+\frac{1}{2\beta_0}\int d^4 x {\sqrt {-q}}e^{-3\phi+\frac{5}{2}\rho}
\;{}^{(4)}\!R^{\mu\nu}\stac{(1)}{T}_{\mu\nu} \nonumber \\
& & -\frac{\beta_0}{8}\int d^4x {\sqrt {-q}}e^{-\phi}
\biggl[ {\rm Tr}(B^4)-\frac{1}{4}\bigl({\rm Tr}(B^2)\bigr)^2\biggr],
\end{eqnarray}
%
where we set $\beta_0=\gamma_0$ so that the Born-Infeld 
action is realised for the flat D-brane with the constant 
field $B_{\mu\nu}$. We also defined 
%
\begin{eqnarray}
\stac{(1)}{T}_{\mu\nu}=-(B^2)_{\mu\nu}+\frac{1}{4}q_{\mu\nu}{\rm Tr}(B^2).
\end{eqnarray}
%

\subsubsection{Summary}

The total solution to the Hamilton-Jacobi equation is summarised by 
%
\begin{eqnarray}
S_{\rm ct}  =  -(S_0+S_1+S_2 +\cdots) = -(\tilde S_{BI} +\tilde S_{EH} +\tilde S_{WZ}
 +\tilde S_2),
\end{eqnarray}
%
where 
%
\begin{eqnarray}
\tilde S_{BI}=  \beta_0 \int d^4 x{\sqrt {-q}}e^{-\phi } \biggl\{ 
1+\frac{1}{4} B_{\mu\nu}B^{\mu\nu}-\frac{1}{8}
\biggl[ {\rm Tr}(B^4) -\frac{1}{4}\bigl({\rm Tr}(B^2)\bigr)^2\biggr] \biggr\},
\end{eqnarray}
%
%
\begin{eqnarray}
\tilde S_{EH} = \frac{1}{\beta_0} \int d^4 x {\sqrt {-q}}
\biggl\{ e^{-3\phi +\frac{5}{2}\rho} \biggl[
\frac{1}{2} {}^{(4)}\!R+4(D\phi )^2+\frac{35}{16}(D\rho )^2 
-\frac{25}{4} D\phi D \rho \biggr] -\frac{1}{4}e^{-\phi +\frac{5}{2}\rho}
(D\chi )^2 \biggr\},
\end{eqnarray}
%
%
\begin{eqnarray}
\tilde S_{WZ} = \beta_0 \int d^4x {\sqrt {-q}} \epsilon^{\mu\nu\alpha\beta}
\biggl[\frac{1}{24}D_{\mu\nu\alpha\beta}+\frac{1}{4}B_{\mu\nu}C_{\alpha\beta}
+\frac{\chi}{8}B_{\mu\nu} B_{\alpha\beta} \biggr],
\end{eqnarray}
%
and 
%
\begin{eqnarray}
\tilde S_{2} = \frac{1}{20\beta_0^3} \int d^4 x {\sqrt {-q}} 
e^{-5 \phi + 5\rho}\;{}^{(4)}\!R_{\mu\nu} {}^{(4)}\!R^{\mu\nu}
+\frac{1}{2\beta_0} \int d^4 x {\sqrt {-q}}
e^{-3\phi +\frac{5}{2}\rho} \;{}^{(4)}\!R^{\mu\nu}\stac{(1)}{T}_{\mu\nu}
+\cdots.
\end{eqnarray}
%
Note that in the flat and constant field limit 
$\tilde S_{BI} =S_{BI}$ up to the current order.
It is also noted that there is non-trivial coupling 
${}^{(4)}\!R^{\mu\nu}\stac{(1)}{T}_{\mu\nu}$ and so on.

\subsection{Effective equation on D-brane}
\label{effective}

\subsubsection{Strategy}

In the braneworld 
the AdS/CFT correspondence may be formulated by the following partition 
functional argument\cite{Giddings}:
%
\begin{eqnarray}
Z & = & \int {\cal D}g e^{iS_{\rm bulk}(g)
+i\frac{1}{2}S_{\rm  D\mbox{\tiny -}brane}(q)+iS_{\rm GH}(q)} \nonumber \\
 & = & \int {\cal D}q e^{\frac{i}{2}S_{\rm D\mbox{\tiny -}brane}+iS_{\rm ct}} \langle 
e^{i\int d^4 x q_{\mu\nu} T^{\mu\nu}}\rangle_{\rm CFT}, 
\end{eqnarray}
%
where $S_{\rm ct}$ represents the counter-terms which make the action finite and 
is given by the on-shell solution of the Hamilton-Jacobi equation,  
$S_{\rm ct}=-(S_0+S_1+S_2 +\cdots )$. 
The variational principle implies 
%
\begin{eqnarray}
\frac{1}{{\sqrt {-q}}}\frac{\delta S_{\rm D\mbox{-}brane}}{\delta q_{\mu\nu}} +2\frac{1}{{\sqrt
{-q}}} 
\biggl( \frac{\delta S_{\rm ct}}{\delta q_{\mu\nu}}
+\frac{\delta \Gamma_{\rm CFT}}{\delta q_{\mu\nu}}  \biggr)=0.\label{principle}
\end{eqnarray}
%
To fix the first term we must specify the D-barne (cut-off brane) action. 
In this paper we consider two types of branes.

\subsubsection{Born-Infeld membrane}

First let us examine the case where the brane action is 
the Born-Infeld type:
%
\begin{eqnarray}
S_{\rm D\mbox{-}brane} & = & \beta \int d^4 x e^{-\phi} 
{\sqrt {-{\rm det}(q_{\mu\nu}+B_{\mu\nu})}}. \label{BIaction}
\end{eqnarray}
%
The energy-momentum tensor of this brane becomes 
%
\begin{eqnarray}
T_{{\rm BI}\mu\nu} = \beta e^{-\phi}q_{\mu\nu} 
-\beta e^{-\phi} \stac{(1)}{T}_{\mu\nu}+\stac{(2)}{T}_{{\rm BI}\mu\nu},
\end{eqnarray}
%
and
%
\begin{eqnarray}
\stac{(2)}{T}_{{\rm BI}\mu\nu}  =  \beta e^{-\phi} \biggl\{ 
-\frac{1}{4}{\rm Tr}(B^2) \biggl[ (B^2)_{\mu\nu}-\frac{1}{8}q_{\mu\nu}
{\rm Tr}(B^2) \biggr] 
+(B^4)_{\mu\nu}-\frac{1}{8}q_{\mu\nu}{\rm Tr}(B^4) \biggr\}.
\end{eqnarray}
%
Substituting Eq.~(\ref{BIaction}) into Eq.~(\ref{principle}), we can 
obtain the effective gravitational equation on the brane. At that time 
we set 
%
\begin{eqnarray}
\beta=2\beta_0,
\end{eqnarray}
%
so that the brane geometry could be four dimensional Minkowski spacetime. 

In the first order the effective equation becomes just vacuum one:
%
\begin{eqnarray}
{}^{(4)}G_{\mu\nu} & = &   
\beta_0 e^{3\phi-\frac{5}{2}\rho} 
\biggl(-\frac{1}{2} \stac{(2)}{T}_{{\rm BI}\mu\nu}+2\frac{1}{{\sqrt {-q}}}
\frac{\delta S_2}{\delta q_{\alpha\beta}} 
q_{\mu\alpha}q_{\nu\beta} \biggr) +T_{\mu\nu}^{\rm CFT} 
\nonumber \\
& = & 
-3(D_\mu D_\nu -q_{\mu\nu}D^2)\phi 
+\frac{5}{2}(D_\mu D_\nu -q_{\mu\nu} D^2)\rho 
+\Bigl[ D_\mu \phi D_\nu \phi -5 q_{\mu\nu}(D\phi )^2 \Bigr] 
 +\frac{15}{8}\biggl[ D_\mu \rho D_\nu \rho -\frac{13}{6}q_{\mu\nu} (D\rho )^2\biggr] \nonumber \\
&  & +\frac{1}{2}e^{2\phi} \biggl[ D_\mu \chi D_\nu \chi -\frac{1}{2}
q_{\mu\nu} (D\chi )^2\biggr] 
-\frac{5}{4}\Bigl( D_\mu \rho D_\nu \phi +D_\mu \phi D_\nu \rho 
-7 q_{\mu\nu} D\phi D\rho\Bigr) 
+T_{\mu\nu}^{\rm CFT}+\cdots. \label{basic1}
\end{eqnarray}
%
We may naively expect that  Einstein-Maxwell theory governs 
the physics on the D-brane described by the Born-Infeld action. 
However, the result is not 
the case. Since $\tilde S_{BI}$ is same as $S_{BI}$ up to the order of 
$(B^4)_{\mu\nu}$, 
the first order Einstein equation does not have the source 
of the Maxwell field while the contribution from  holographic CFT exists. 

As a result, the gravitational equation up to the second order is given by
%
\begin{eqnarray}
{}^{(4)}G_{\mu\nu}  
& = & T_{\mu\nu}^{\rm CFT}
-3\bigl(D_\mu D_\nu -q_{\mu\nu}D^2\bigr)\phi 
+\frac{5}{2}\bigl(D_\mu D_\nu -q_{\mu\nu} D^2\bigr)\rho 
+\biggl[ D_\mu \phi D_\nu \phi -5 q_{\mu\nu}(D\phi )^2 \biggr] \nonumber \\
& &  +\frac{15}{8}\biggl[ D_\mu \rho D_\nu \rho -\frac{13}{6}g_{\mu\nu} (D\rho )^2\biggr] 
+\frac{1}{4}e^{2\phi} \biggl[ D_\mu \chi D_\nu \chi -\frac{1}{2}
q_{\mu\nu} (D\chi )^2\biggr] \nonumber \\
& & -\frac{5}{4}\Bigl( D_\mu \rho D_\nu \phi +D_\mu \phi D_\nu \rho 
-7 q_{\mu\nu} D\phi D\rho\Bigr) \nonumber \\
& & -\frac{1}{5\beta_0^2}e^{-2\phi +\frac{5}{2}\rho}
\biggl({}^{(4)}\!R_{\mu}^{~\alpha}{}^{(4)}\!R_{\alpha\nu}
-\frac{1}{4}q_{\mu\nu}{}^{(4)}\!R_{\alpha\beta}R^{\alpha\beta} \biggr)
-\frac{1}{10\beta_0^2} e^{-2\phi+\frac{5}{2}\rho}D^2 \;{}^{(4)}\!R_{\mu\nu}
\nonumber \\
& & +{}^{(4)}\!R_{\mu\alpha}(B^2)^\alpha_{~\nu}
+{}^{(4)}\!R_{\nu\alpha} (B^2)^\alpha_{~\mu} 
-\frac{1}{4}{}^{(4)}\!R_{\mu\nu}{\rm Tr}(B^2)
+{}^{(4)}\!R^{\alpha\beta} B_{\mu\alpha} B_{\beta\nu}
-\frac{1}{2}q_{\mu\nu}{}^{(4)}\!R_{\alpha\beta}(B^2)^{\alpha\beta}
-\frac{1}{2}D^2 \stac{(1)}T_{\mu\nu}.
\end{eqnarray}
%
In the above we have dropped the second order terms which 
couple to the scalar fields to keep the form compact.

\subsubsection{Nambu-Goto membrane}

For the comparison, it might be worth considering the brane described by 
the Nambu-Goto action
%
\begin{eqnarray}
S_{\rm NG}=2\beta_0 \int d^4 x{\sqrt {-q}}e^{-\phi}.
\end{eqnarray}
%
At the first order, the effective equation becomes 
%
\begin{eqnarray}
{}^{(4)}G_{\mu\nu} & = & \beta_0^2 e^{2\phi-\frac{5}{2}\rho} 
\stac{(1)}{T}_{\mu\nu}
-3\bigl(D_\mu D_\nu -q_{\mu\nu}D^2\bigr)\phi 
+\frac{5}{2}\bigl(D_\mu D_\nu -q_{\mu\nu} D^2\bigr)\rho \nonumber \\
& & +\biggl[ D_\mu \phi D_\nu \phi -5 q_{\mu\nu}(D\phi )^2 \biggr]
+\frac{15}{8}\biggl[ D_\mu \rho D_\nu \rho -\frac{13}{6}q_{\mu\nu} (D\rho )^2\biggr]
+\frac{1}{4}e^{2\phi} \biggl[ D_\mu \chi D_\nu \chi -\frac{1}{2}
g_{\mu\nu} (D\chi )^2\biggr] \nonumber \\
& & -\frac{5}{4}\Bigl( D_\mu \rho D_\nu \phi +D_\mu \phi D_\nu \rho 
-7 q_{\mu\nu} D_{\alpha}\phi D^{\alpha}\rho\Bigr)+T_{\mu\nu}^{\rm CFT}+\cdots.
\end{eqnarray}
%
The cancellation does not occur and  Einstein-Maxwell-scalar theory 
is realised on the brane. This is also an unexpected result.

\section{Geometrical approach}

In the previous section, we saw unexpected results for the effective theory on the brane. 
It seems that they are originated from the fact that the Born-Infeld action is a solution 
to the Hamilton-Jacobi equation. In order to understand why we obtain such consequences, 
we will re-derive the gravitational equation on the D-brane using the 
geometrical method\cite{Tess,GE} in this section. To do so we will solve the bulk spacetime in
the  long wave approximation and then  obtain the effective theory in the low-energy limit. 
We will use the slightly different notations from the previous sections. 

The rest of this section is organised as follows. In the Subsec.~\ref{subsecIIIA}, we 
give a formulation of the geometrical approach and  stress that we must 
solve the bulk fields and gravity somehow. Then we solve them in long wave 
approximation up to leading order for the gravitational theory on the brane. 
Finally we derive the gravitational equation on the D-branes described by 
Born-Infeld action in Subsec.~\ref{subsecIIIB}.

\subsection{formulation}
\label{subsecIIIA}

The full metric is written as
%
\begin{eqnarray}
ds^2=e^{2\varphi (x)}dy^2+q_{\mu\nu}(y,x) dx^\mu dx^\nu.
\end{eqnarray}
%
The induced metric on the brane is $h_{\mu\nu}(x)=q_{\mu\nu}(y_0,x)$, where we suppose that the 
brane is located at $y=y_0$. 

In the geometrical approach, the gravitational equation on the brane is given by 
%
\begin{eqnarray}
{}^{(4)}G_{\mu\nu}(h) =  \frac{2}{3}\biggl[T_{\mu\nu}+h_{\mu\nu}
\Bigl(T_{yy}-\frac{1}{4}T \Bigr) \biggr]  
+KK_{\mu\nu}-K_\mu^{~\alpha} K_{\nu\alpha} 
 -\frac{1}{2}(K^2-K_{\alpha\beta}K^{\alpha\beta})h_{\mu\nu}-E_{\mu\nu},
\label{SMS}
\end{eqnarray}
%
where 
%
\begin{eqnarray}
T_{MN} & = & -2 \bigl(\nabla_M \nabla_N -g_{MN}\nabla^2 \bigr)\phi 
+\frac{5}{4}\bigl(\nabla_M \nabla_N -g_{MN}\nabla^2\bigr)\rho 
+\frac{1}{2}e^{2\phi}
\biggl[ \nabla_M \chi \nabla_N \chi -\frac{1}{2}g_{MN} (\nabla \chi )^2\biggr] 
\nonumber \\
& & +\frac{5}{16}\biggl[ \nabla_M \rho \nabla_N \rho -3 g_{MN}(\nabla \rho )^2\biggr]
-2g_{MN}(\nabla \phi )^2 +\frac{5}{2}g_{MN} \nabla_K \phi \nabla^K \rho 
+\frac{1}{4}\bigl(H_{MKL}H_N^{~~KL}-g_{MN}|H|^2 \bigr) \nonumber \\
& & +\frac{1}{4}e^{2\phi} \bigl(\tilde F_{MKL}\tilde F_N^{~~KL}-g_{MN}|\tilde F|^2\bigr)
+\frac{1}{96}e^{2\phi}\tilde G_{MK_1 K_2 K_3 K_4} \tilde G_N^{~~K_1 K_2 K_3 K_4},
\end{eqnarray}
%
and then 
%
\begin{eqnarray}
T_{\mu\nu}+h_{\mu\nu}\Bigl(T_{yy}-\frac{1}{4}T \Bigr)
& = & -2\bigl(D_\mu D_\nu \phi-h_{\mu\nu}D^2\phi\bigr)
+\frac{5}{4}\bigl(D_\mu D_\nu \rho -h_{\mu\nu}D^2\rho\bigr) 
+\frac{1}{2}e^{2\phi} \biggl[D_\mu \chi D_\nu \chi 
-\frac{5}{8}h_{\mu\nu}(D\chi)^2 \biggr] \nonumber \\
& & +\frac{5}{16}\biggl[D_\mu \rho D_\nu \rho -\frac{5}{2}h_{\mu\nu}(D\rho )^2 
\biggr] 
-\frac{3}{2}h_{\mu\nu}(D\phi)^2+\frac{15}{8}h_{\mu\nu}D_{\alpha}\phi D^{\alpha}\rho \nonumber \\
& & -2(K_{\mu\nu}-h_{\mu\nu}K)\partial_y \phi 
+\frac{5}{4}(K_{\mu\nu}-h_{\mu\nu}K)\partial_y \rho \nonumber \\
& & +\frac{3}{16}e^{2\phi}h_{\mu\nu}(\partial_y \chi)^2
-\frac{15}{32}h_{\mu\nu}(\partial_y \rho)^2
-\frac{3}{2}h_{\mu\nu} (\partial_y \phi)^2 
+\frac{15}{8}h_{\mu\nu}\partial_y \phi \partial_y \rho \nonumber \\
& & +\frac{1}{2}\biggl(H_{y\mu\alpha}H_{y\nu}^{~~\alpha}
-\frac{1}{16}h_{\mu\nu}H_{y\alpha\beta}H^{y\alpha\beta} \biggr) 
+\frac{1}{2}e^{2\phi}
\biggl(\tilde F_{y\mu\alpha} \tilde F_{y\nu}^{~~\alpha}
-\frac{1}{16}h_{\mu\nu} \tilde F_{y\alpha\beta}\tilde F^{y\alpha\beta} \biggr) \nonumber \\
& & +\frac{1}{24}e^{2\phi}\biggl( \tilde G_{\mu y \alpha_1 \alpha_2 \alpha_3}
\tilde G_{\nu y}^{~~~\alpha_1 \alpha_2 \alpha_3} 
-\frac{1}{16}h_{\mu\nu} \tilde G_{y \alpha_1 \alpha_2 \alpha_3 \alpha_4}
\tilde G^{y\alpha_1 \alpha_2 \alpha_3 \alpha_4} \biggr). 
\end{eqnarray}
%
$E_{\mu\nu}$ is the projected five-dimensional Weyl tensor defined by 
$E_{\mu\nu}={}^{(5)}C_{\mu M \nu N}n^M n^N$. 
It is obvious that the above equation is not closed in four dimensions. Moreover, 
when bulk fields exist, $E_{\mu\nu}$ is not negligible in low energy
limits\cite{holo3,Hime,GE2}.  Since the Born-Infeld action appears as a solution to the
Hamilton-Jacobi equation,   we guess that $E_{\mu\nu}$ contains a part of the Born-Infeld
energy-momentum tensor. 

As the previous section, for simplicity, we turn off almost fields 
except scalar fields, $B_{\mu\nu}$ and $\tilde G_{y \mu_1 \mu_2 \mu_3 \mu_4}$. 

To obtain the background solution which is consistent with the junction condition, 
we assume that the action for the brane is given by 
%
\begin{eqnarray}
S_{\rm brane}=2\beta \int d^4 xe^{-\phi}{\sqrt {-{\rm det}(h+B)}}
+2 \beta \int d^4x {\sqrt {-h}}\epsilon^{\mu\nu\alpha\beta} \biggl[ 
\frac{1}{4}B_{\mu\nu}C_{\alpha\beta}+\frac{\chi}{8}B_{\mu\nu}B_{\alpha\beta}
+\frac{1}{24}D_{\mu\nu\alpha\beta}
\biggr].
\label{brane}
\end{eqnarray}
%

The boundary conditions at the brane are brought by the junction conditions:
%
\begin{eqnarray}
\biggl[ (K_{\mu\nu}-h_{\mu\nu}K)e^\varphi 
+\biggl( 2 \partial_y \phi -\frac{5}{4}\partial_y \rho \biggr)h_{\mu\nu}
\biggr](y_0,x)=-\frac{1}{4} e^{\varphi + \phi-\frac{5}{4}\rho} T_{\mu\nu}^{\rm BI}, 
\label{jun1}
\end{eqnarray}
%
%
\begin{eqnarray}
\biggl[4K e^\varphi -8\partial_y \phi +5\partial_y \rho \biggr](y_0,x)
= \beta e^{\varphi + \phi -\frac{5}{4}\rho}
\biggl[ 1-\frac{1}{4}{\rm Tr}(B^2) +\frac{1}{32} \bigl({\rm Tr}(B^2) \bigr)^2
-\frac{1}{8}{\rm Tr}(B^4) +O(B^6) \biggr],
\end{eqnarray}
%
%
\begin{eqnarray}
\biggl[-K e^\varphi-\partial_y \rho +2\partial_y \phi \biggr](y_0,x)=0,
\end{eqnarray}
%
%
\begin{eqnarray}
\partial_y \chi (y_0,x) = -\frac{1}{8}\beta e^{\varphi-\frac{5}{4}\rho}\epsilon^{\mu\nu\alpha\beta} 
B_{\mu\nu}B_{\alpha\beta},
\end{eqnarray}
%
%
\begin{eqnarray}
H_{y\mu\nu}(y_0,x)  = - \beta e^{\varphi+ \phi -\frac{5}{4}\rho} 
\biggl[ B_{\mu\nu} -\frac{1}{4}{\rm Tr}(B^2) B_{\mu\nu}
+(B^3)_{\mu\nu} +O(B^5) \biggr], 
\end{eqnarray}
%
%
\begin{eqnarray}
\tilde F_{y\mu\nu} (y_0,x)=-\frac{\beta}{2}e^{\varphi -\frac{5}{4}\rho} 
\epsilon_{\mu\nu\alpha\beta}B^{\alpha\beta},
\end{eqnarray}
%
and
%
\begin{eqnarray}
\tilde G_{y\mu_1 \mu_2 \mu_3 \mu_4} (y_0,x)
=-\beta  e^{\varphi -\frac{5}{4}\rho}\epsilon_{\mu_1 \mu_2 \mu_3 \mu_4},
\label{jun2}
\end{eqnarray}
%
where 
%
\begin{eqnarray}
T_{\mu\nu}^{\rm BI}=2 \beta e^{-\phi}\biggl\{ h_{\mu\nu} -\stac{(1)}{T}_{\mu\nu}
-\frac{1}{4}{\rm Tr}(B^2) \Bigl[ (B^2)_{\mu\nu}-\frac{1}{8}h_{\mu\nu} {\rm Tr}(B^2) \Bigr] 
+(B^4)_{\mu\nu}-\frac{1}{8}h_{\mu\nu}{\rm Tr}(B^4) +O(B^6) \biggr\}.
\end{eqnarray}
%
Using these junction conditions, the Eq.~(\ref{SMS}) becomes 
%
\begin{eqnarray}
{}^{(4)} G_{\mu\nu} (h) & = & \frac{5}{6}\beta^2 e^{2 \phi - \frac{5}{2} \rho} 
\stac{(1)} T_{\mu\nu} 
 - \frac{4}{3} \Bigl( D_{\mu} D_{\nu} - h_{\mu\nu} D^2 \Bigr) \phi 
+ \frac{5}{6} \Bigl( D_{\mu} D_{\nu} - h_{\mu\nu} D^2 \Bigr)\rho
\nonumber\\
& & + \frac{1}{3} e^{2 \phi} \biggl[ D_{\mu} \chi D_{\nu} \chi - \frac{5}{8} h_{\mu\nu} ( D \chi )^2
\biggr] + \frac{5}{24} \biggl[ D_{\mu} \rho D_{\nu} \rho - \frac{5}{2} h_{\mu\nu} ( D \rho )^2
\biggr]
 - h_{\mu\nu} ( D \phi )^2 + \frac{5}{4} h_{\mu\nu} D_{\alpha} \phi D^{\alpha} \rho 
\nonumber\\
& &- E_{\mu\nu}  + O(B^4) . \label{SMS2}
\end{eqnarray} 
%

The bulk ``evolutional" equations are 
%
\begin{eqnarray}
\label{evo1}
\mbox \pounds_n K = {}^{(4)}R-  \biggl(T^\mu_{~\mu} -\frac{4}{3}T 
\biggr) -K^2 -D^2 \varphi -(D\varphi)^2,
\end{eqnarray}
%
%
\begin{eqnarray}
\mbox \pounds_n \tilde K^\mu_{~\nu} ={}^{(4)}\!\tilde R^\mu_{~\nu}
-\bigl(D^{\mu} D_{\nu} \varphi +D^{\mu} \varphi D_{\nu} \varphi\bigr)_{\rm traceless} 
-\biggl(  T^\mu_{~\nu} -\frac{1}{4}\delta^\mu_{~\nu} T^\alpha_{~\alpha} \biggr)  -K \tilde
K^\mu_{~\nu}  ,
\end{eqnarray}
%
%
\begin{eqnarray}
& & 4\partial_y^2 \phi -\frac{5}{2}\partial_y^2 \rho -8 (\partial_y \phi)^2
-\frac{25}{8}(\partial_y \rho)^2+10 \partial_y \phi \partial_y \rho 
-\frac{3}{2}e^{2\phi}(\partial_y \chi)^2 
+e^\varphi K \biggl( 4 \partial_y \phi -\frac{5}{2}\partial_y \rho \biggr) \nonumber \\
& & ~~
+\biggl[
\frac{5}{2}e^{2\phi}|\tilde G|^2 
-D_\mu \varphi D^\mu \biggl(4 \phi -\frac{5}{3}\rho \biggr) 
+4D^2 \phi -\frac{5}{2}D^2\rho -8 (D\phi)^2-\frac{25}{8}(D\rho)^2 \nonumber \\
& & ~~~~~~~~+10D_\alpha \phi D^\alpha \rho -\frac{3}{2}e^{2\phi}(D\chi)^2 
+2|H|^2+\frac{1}{2}e^{2\phi}|\tilde F|^2\biggr]e^{2\varphi} =0,
\end{eqnarray}
%
%
\begin{eqnarray}
& & -\partial_y^2 \phi +\partial_y^2 \rho 
+2(\partial_y \phi)^2+\frac{5}{4}(\partial_y \rho)^2
- \frac{13}{4} \partial_y \phi \partial_y \rho 
-e^\varphi K(\partial_y \phi -\partial_y \rho) 
\nonumber \\
& & ~~
+\biggl[ -e^{2\phi}|\tilde G|^2
+D_\mu \varphi D^\mu (\phi -\rho)
-D^2 \phi +D^2 \rho +2(D\phi)^2+\frac{5}{4}(D\rho)^2
-\frac{13}{4} D_\alpha \phi D^\alpha \rho -\frac{1}{2}|H|^2
-\frac{1}{2}e^{2\phi}|\tilde F|^2\biggr]e^{2\varphi} =0,
\nonumber \\
\end{eqnarray}
%
%
\begin{eqnarray}
\partial_y \bigl(e^{\frac{5}{4}\rho} \partial_y \chi\bigr)+D_\alpha 
\bigl(e^{\frac{5}{4}\rho} D^\alpha \chi\bigr) +e^{\frac{5}{4}\rho} e^\varphi K \partial_y \chi
-e^{2\varphi + \frac{5}{4}\rho}D_\mu \chi D^\mu \varphi=0,
\end{eqnarray}
%
%
\begin{eqnarray}
& & \partial_y \bigl(e^{-2\phi+\frac{5}{4}\rho}H^{y\mu\nu}
+e^{\frac{5}{4}\rho}\chi \tilde F^{y\mu\nu} \bigr) +
e^\varphi K\bigl(e^{-2\phi+\frac{5}{4}\rho} H^{y\mu\nu} +e^{\frac{5}{4}\rho}\tilde F^{y\mu\nu}\bigr) 
+\frac{1}{2}e^{\frac{5}{4}\rho}F_{y\alpha\beta} \tilde G^{\mu\nu y \alpha\beta}=0,
\end{eqnarray}
%
%
\begin{eqnarray}
\partial_y \bigl(e^{\frac{5}{4}\rho}\tilde F^{y\mu\nu}\bigr)
+e^{\varphi + \frac{5}{4}\rho}K \tilde F^{y\mu\nu}-\frac{1}{2}e^{\frac{5}{4}\rho}
H_{y\alpha\beta}\tilde G^{\mu\nu y \alpha\beta}=0,
\end{eqnarray}
%
and
%
\begin{eqnarray}
\label{evo2}
\partial_y \bigl(e^{\frac{5}{4}\rho}\tilde G_{y\mu_1 \mu_2 \mu_3 \mu_4}\bigr)
=K e^{\frac{5}{4}\rho}\tilde G_{y\mu_1 \mu_2 \mu_3 \mu_4}.
\end{eqnarray}
%
The Hamiltonian and momentum constraints are 
%
\begin{eqnarray}
-\frac{1}{2}\biggl[{}^{(4)}\!R-\frac{3}{4}K^2 +\tilde K^\alpha_{~\beta} 
\tilde K^\beta_{~\alpha} \biggr]= T_{yy}e^{-2\varphi},
\end{eqnarray}
%
and
%
\begin{eqnarray}
D_\nu K^\nu_{~\mu} -D_\mu K =  T_{\mu y}e^{-\varphi},
\end{eqnarray}
%
respectively. The constraints for $H_{y\mu\nu}$, $\tilde F_{y\mu\nu}$ and 
$\tilde G_{y\mu_1 \mu_2 \mu_3 \mu_4}$ are 
%
\begin{eqnarray}
D^\alpha \bigl(e^{-\varphi-2\phi+\frac{5}{4}\rho}H_{y\alpha\mu}+e^{-\varphi+\frac{5}{4}\rho}
\chi \tilde F_{y\alpha\mu} \bigr)=0,
\end{eqnarray}
%
%
\begin{eqnarray}
D^\alpha \bigl(e^{\varphi+\frac{5}{4}\rho}\tilde F_{y\alpha\mu} \bigr)=0,
\end{eqnarray}
%
and
%
\begin{eqnarray}
D^\alpha \bigl(e^{-\varphi+\frac{5}{4}\rho}\tilde G_{y \alpha \mu_1 \mu_2 \mu_3}\bigr)=0.
\end{eqnarray}
%

\subsection{Solving of the bulk and effective theory}
\label{subsecIIIB}

Let us solve the bulk equations (\ref{evo1})-(\ref{evo2}) with the junction conditions
(\ref{jun1})-(\ref{jun2}) in the long wave approximation. The  infinitesimal parameter of the
expansion is $\epsilon=(\ell /L)^2$, where
$\ell$ is the bulk  curvature scale and $L$ is the typical scale on the brane.

\subsubsection{0th-order}
\label{IIIB-1}

In the 0th order the evolutional equations are 
%
\begin{eqnarray}
e^{-\varphi}\partial_y \stac{(0)}{\tilde K^\mu_{~\nu}}=
-\stac{(0)}{K}\stac{(0)}{\tilde K^\mu_{~\nu}},
\end{eqnarray}
%
%
\begin{eqnarray}
e^{-\varphi}\partial_y \stac{(0)}{K} =-\biggl(T^\mu_{~\mu} - \frac{4}{3}T 
\biggr)^{(0)}-\stac{(0)}{K^2},
\end{eqnarray}
%
%
\begin{eqnarray}
4 \partial_y^2 \phi_0 -\frac{5}{2}\partial_y^2 \rho_0 
+e^\varphi \stac{(0)}{K} \biggl( 4\partial_y \phi_0 - \frac{5}{2}\partial_y 
\rho_0\biggr) -8 (\partial_y \phi_0)^2 -\frac{25}{8} (\partial_y \rho_0)^2
+10 \partial_y \phi_0 \partial_y \rho_0 
+\frac{5}{2}e^{2\varphi + 2\phi_0} |\tilde G|^2=0,
\end{eqnarray}
%
%
\begin{eqnarray}
 -\partial_y^2 \phi_0 + \partial_y^2 \rho_0 +e^\varphi \stac{(0)}{K}(-\partial_y \phi_0 
+\partial_y \rho_0 ) +2(\partial_y \phi_0)^2+\frac{5}{4}(\partial_y \rho_0)^2
-\frac{13}{4}\partial_y \phi_0 \partial_y \rho_0 
-e^{2\varphi+ 2\phi_0} |\tilde G|^2
=0,
\end{eqnarray}
%
and
%
\begin{eqnarray}
\partial_y \bigl(e^{\frac{5}{4}\rho_0}\tilde G_{y \mu_1 \mu_2 \mu_3 \mu_4} \bigr)
=\stac{(0)}{K} e^{\frac{5}{4}\rho_0} \tilde G_{y \mu_1 \mu_2 \mu_3 \mu_4}.
\end{eqnarray}
%
The constraint equations are
%
\begin{eqnarray}
D^{\mu_1} \bigl(e^{-\varphi+\frac{5}{4}\rho_0}\tilde G_{y \mu_1 \mu_2 \mu_3 \mu_4} \bigr)=0.
\end{eqnarray}
%
The junction conditions are 
%
\begin{eqnarray}
\stac{(0)}{{\tilde K}^\mu_{~\nu}}(y_0,x)=0,
~~~
\stac{(0)}{K}(y_0,x)=- \beta e^{\phi_0 -\frac{5}{4}\rho_0},
~~~
\partial_y \phi_0 (y_0,x)=0,
~~~
\partial_y \rho_0 (y_0,x)=\beta e^{\varphi+ \phi_0-\frac{5}{4}\rho_0},
\end{eqnarray}
%
and
%
\begin{eqnarray}
\tilde G_{y \mu_1 \mu_2 \mu_3 \mu_4}(y_0,x)
=- \beta e^{\varphi -\frac{5}{4}\rho_0}\epsilon_{\mu_1 \mu_2 \mu_3 \mu_4}.
\end{eqnarray}
%

The background solution is easily found as\footnote{There could be another solution 
which cannot be written in an analytic form. But, we cannot derive the 
effective equation in an analytic form at leading order.}
%
\begin{eqnarray}
&&\phi_0 =\phi_0 (x),
~~~
\rho_0=\frac{4}{5}{\rm log}(y/y_0)+\sigma_0(x), 
~~~
\chi_0=\chi_0(x),
\nonumber \\
&&\stac{(0)}C_{\mu\nu}=\stac{(0)}{B}_{\mu\nu}=0,
~~~
{\rm and}
~~~
\tilde G_{y \mu_1 \mu_2 \mu_3 \mu_4}=-\beta e^{\varphi-\frac{5}{4}\rho_0}
\epsilon_{\mu_1 \mu_2 \mu_3 \mu_4},
\end{eqnarray}
%
where 
%
\begin{equation}
\varphi (x)=-\phi_0(x)+\frac{5}{4}\sigma_0 (x) +{\rm log}(4/5y_0 \beta).
\end{equation}
%
The extrinsic curvature is given by 
%
\begin{eqnarray}
\stac{(0)}{K^\mu_{~\nu}}=-\frac{1}{5y}e^{-\varphi (x)} \delta^\mu_{~\nu},
\end{eqnarray}
%
and then the metric becomes 
%
\begin{eqnarray}
\stac{(0)}g_{\mu\nu}=a^2(y) h_{\mu\nu}(x),
\end{eqnarray}
%
where 
%
\begin{eqnarray}
a(y)= \biggl( \frac{y}{y_0} \biggr)^{-\frac{1}{5}}.
\end{eqnarray}
%
The behavior of the background metric brings us a serious problem, that is, 
the four-dimensional gravity cannot be recovered on the brane.  The minimum way to 
see this is the dimensional reduction from five to four dimensions
%
\begin{eqnarray}
\int d^5x {\sqrt {-g}}\;{}^{(5)}\!R \sim \int dy a^2(y) \int d^4x {\sqrt {-h}}\;{}^{(4)}\!R(h).
\end{eqnarray}
%
In the above $\int dy a^2(y) = \infty$ when we consider the infinite extra dimensions,
which implies that
the four dimensional gravity cannot be recovered.
This problem might be regarded as a sort of no-go theorem proposed by 
Maldacena\cite{nogo}. The simple resolution to this problem is the compactification and/or
introduction of the another brane. There may be another possibility that the bulk 
or brane action is modified via some quantum effects. We leave this issue for future study. 

Note that adding a Wess-Zumino term  $\int D$ in the brane action
is essential to obtain solutions. There is no solution with 
similar form when brane is supposed to be described only by Nambu-Goto
action. This fact is consistent with that the Nambu-Goto action alone 
cannot satisfy the Hamilton-Jacobi equation.

\subsubsection{1st order}

Next we turn to the  1st order equations. The junction conditions become
%
\begin{eqnarray}
\stac{(1)}{K}(y_0,x)=-\frac{1}{4}\beta e^{\phi_0-\frac{5}{4}\rho_0}{\rm Tr}(B^2),
\end{eqnarray}
%
%
\begin{eqnarray}
\stac{(1)}{{\tilde K}^\mu_{~\nu}}(y_0,x)
=\frac{1}{2}\beta e^{\phi_0-\frac{5}{4}\rho_0}\stac{(1)}{T^\mu_{~\nu}},
\end{eqnarray}
%
%
\begin{eqnarray}
\partial_y \phi_1 (y_0,x)=-\frac{1}{4}\beta e^{\varphi + \phi_0-\frac{5}{4}\rho_0}{\rm Tr}(B^2),
\end{eqnarray}
%
and
%
\begin{eqnarray}
\partial_y \rho_1 (y_0,x) = 
-\frac{1}{4} \beta e^{\varphi + \phi_0-\frac{5}{4}\rho_0}{\rm Tr}(B^2).
\end{eqnarray}
%

For the gravitational equation on the brane, the key equations are the 
evolutional equation for the traceless part of the extrinsic curvature 
and the Hamiltonian constraint:
%
\begin{eqnarray}
\partial_y \stac{(1)}{{\tilde K}^\mu_{~\nu}}= {}^{(4)} \!\tilde R^\mu_{~\nu} 
-(D^{\mu} D_{\nu} \varphi +D^{\mu} \varphi D_{\nu} \varphi )_{\rm traceless} 
-\biggl(T^\mu_{~\nu} -\frac{1}{4}\delta^\mu_{~\nu} T^\alpha_{~\alpha} 
\biggr)^{(1)}-\stac{(0)}{K} \stac{(1)}{{\tilde K}^\mu_{~\nu}} ,
\label{traceless}
\end{eqnarray}
%
and
%
\begin{eqnarray}
-\frac{1}{2}{}^{(4)}\!R+\frac{3}{4}\stac{(0)}{K} \stac{(1)}{K}
= \stac{(1)}{T_{yy}}e^{-2\varphi}.
\end{eqnarray}
%
Since the right-hand side in Eq.~(\ref{traceless}) 
contains $\tilde F_{y\mu\nu}$ and $H_{y\mu\nu}$, we also need to solve 
their bulk equations
%
\begin{eqnarray}
\partial_y X^{\mu\nu} +e^\varphi \stac{(0)}{K} X^{\mu\nu}+\frac{1}{2}e^{\frac{5}{4}\rho_0}F_{y\alpha\beta}
\tilde G^{y\alpha\beta\mu\nu}=0, 
\label{Heq}
\end{eqnarray}
%
and
%
\begin{eqnarray}
\partial_y (e^{\frac{5}{4}\rho_0}\tilde F^{y\mu\nu})+e^{\varphi+\frac{5}{4}\rho_0}
\stac{(0)}{K} \tilde F^{y\mu\nu}-\frac{1}{2}e^{\frac{5}{4}\rho_0} H_{y\alpha\beta}
\tilde G^{y\alpha\beta\mu\nu}=0, 
\label{Feq}
\end{eqnarray}
%
where $X^{\mu\nu}=e^{-2\phi_0+\frac{5}{4}\rho_0}H^{y\mu\nu}
+e^{\frac{5}{4}\rho_0} \chi \tilde F^{y\mu\nu}$ and $ (T^\mu_\nu 
-\frac{1}{4}\delta^\mu_\nu T^\alpha_\alpha)^{(1)}$ is the 
first order part of $(T^\mu_\nu 
-\frac{1}{4}\delta^\mu_\nu T^\alpha_\alpha)$. The solutions are easily found 
%
\begin{eqnarray}
H_{y\mu\nu}(y,x)=-
a(y)^9\beta e^{\varphi+\phi_0-\frac{5}{4}\sigma_0}B_{\mu\nu}(x),
\end{eqnarray}
%
and
%
\begin{eqnarray}
\tilde F_{y\mu\nu} (y,x)=-a(y)^9\frac{\beta}{2}
e^{\varphi-\frac{5}{4}\sigma_0} \hat \epsilon_{\mu\nu\rho\sigma}(h) B_{\alpha\beta}h^{\rho\alpha}
h^{\sigma\beta}.
\end{eqnarray}
%

Let us derive the gravitational equation on the brane. 
{}From the Hamiltonian constraint on the brane, we first obtain 
%
\begin{eqnarray}
{}^{(4)}\!\hat R (h)=-2\hat D^2 \phi_0 +\frac{6}{5}(\hat D\phi_0)^2+\frac{1}{2}e^{2\phi_0}
(\hat D\chi_0)^2,
\end{eqnarray}
%
where $\hat D_\mu$ is the covariant derivative with respect to $h_{\mu\nu}$. 

Substituting the above solutions into Eq.~(\ref{traceless}) and integrating 
over $y$, we obtain 
%
\begin{eqnarray}
\frac{1}{y_0}\stac{(1)}{{\tilde K}^\mu_{~\nu}}(y,x) & = & \frac{5}{8}a^{-7}
e^\varphi \biggl[{}^{(4)}\!\hat R^\mu_\nu (h)+2 \hat D^\mu \hat D_\nu 
\phi_0 -\frac{5}{4}\hat D^\mu \hat D_\nu \rho_0 
-\frac{5}{16}\hat D^\mu \rho_0 \hat D_\nu \rho_0
-\frac{5}{4}\hat D^\mu \rho_0 \hat D_\nu \phi_0 \nonumber\\
& & -\frac{5}{4}\hat D^\mu \phi_0 \hat D_\nu \rho_0
 -\frac{1}{2}e^{2\phi_0}\hat D^\mu \chi_0 
\hat D_\nu \chi_0 -\hat D^\mu \hat D_\nu \varphi -\hat D^\mu \varphi 
\hat D_\nu \varphi 
-a^{14} \beta^2 e^{\varphi + 2\phi_0-\frac{5}{2}\sigma_0} \hat B_\mu^{~\alpha}\hat B_{\nu\alpha}
\biggr]_{\rm traceless} \nonumber \\
& & +a \chi^\mu_{~\nu} (x),
\end{eqnarray}
%
where $\chi_{\mu\nu}(x)$ is the constant of integration. Together with the junction condition for 
$K^\mu_{~\nu}$ and the Hamiltonian constraint, we finally
obtain the effective equation on
the brane:
%
\begin{eqnarray}
{}^{(4)}G_{\mu\nu} (h) & = & 
-\biggl( \hat D_\mu \hat D_\nu +\frac{3}{4}h_{\mu\nu}\hat D^2 \biggr)
\biggl( 2\phi_0 -\frac{5}{4}\rho_0 \biggr)
+\frac{16}{5}\biggl( \hat D_\mu \rho_0 \hat D_\nu \rho_0 -\frac{7}{4}h_{\mu\nu}(\hat D \rho_0)^2 \biggr) 
\nonumber \\
& & 
+\frac{1}{2}e^{2\phi_0}\biggl[ \hat D_\mu \chi_0 \hat D_\nu \chi_0 
-\frac{1}{2}h_{\mu\nu}(\hat D
\chi_0)^2 \biggr] 
-(\hat D \phi_0)^2 h_{\mu\nu}+
\frac{5}{4}\hat D_\alpha \phi_0 \hat D^\alpha \rho_0 h_{\mu\nu} \nonumber \\
& & +
\hat D_\mu \hat D_\nu \varphi -\frac{1}{4}h_{\mu\nu}(\hat D \varphi )^2
+\hat D_\mu \varphi \hat D_\nu \varphi -\frac{1}{4}h_{\mu\nu}(\hat D\varphi)^2
+\tilde \chi_{\mu\nu}(x).
\label{basic3}
\end{eqnarray}
%
This is main result in this section. 
Although we can write $\varphi$ in terms of $\phi_0$ and $\rho_0$, 
we leave it from a pedagogical point of view. 
As in the previous section, it turns out again 
that the gravitational equation on the brane is {\it not} like Einstein-Maxwell 
theory. $\stac{(1)}{T}_{\mu\nu}$ is exactly canceled out! 

Comparing Eq.~(\ref{basic3}) with (\ref{SMS2}), we find that the relation between
$E_{\mu\nu}$ and 
$\tilde \chi_{\mu\nu}$ is 
%
\begin{eqnarray}
-E_{\mu\nu}& = & \tilde \chi_{\mu\nu}-\frac{5}{6}\beta^2 e^{2\phi_0 -\frac{5}{2}\rho_0}
\stac{(1)}{T}_{\mu\nu} 
-\frac{5}{3} \biggl[ \hat D_\mu \hat D_\nu \phi_0 -\frac{1}{4}h_{\mu\nu}(\hat D\phi_0)^2 \biggr]
+\frac{5}{3} \biggl[ \hat D_\mu \hat D_\nu \rho_0 -\frac{1}{4}h_{\mu\nu} (\hat D\rho_0)^2\biggr] \nonumber
\\ & & +\frac{1}{6}e^{2\phi}\biggl[ \hat D_\mu \chi_0 \hat D_\nu \chi_0 -\frac{1}{4}h_{\mu\nu}(\hat D
\chi_0)^2\biggr] +\frac{5}{3}\biggl[ \hat D_\mu \rho_0 \hat D_\nu \rho_0 -\frac{1}{4}h_{\mu\nu}(\hat D
\rho_0)^2 \biggr] +\hat D_\mu \phi_0 \hat D_\nu \phi_0 -\frac{1}{4}h_{\mu\nu}(\hat D\phi_0)^2 \nonumber \\
& & 
-\frac{5}{4}\biggl[ \hat D_\mu \phi_0 \hat D_\nu \rho_0+\hat D_\mu \rho_0 \hat D_\nu \phi_0
-\frac{1}{2}h_{\mu\nu} \hat D_\mu  \rho_0 \hat D^\mu \phi_0 \biggr].
\end{eqnarray}
%

We should again notice that the form of the brane action Eq.~(\ref{brane}) 
is essential to have consistent solutions. Solutions for 
$H_{y \mu \nu}$ and $\tilde{F}_{y \mu \nu}$ in the bulk 
are automatically consistent with boundary conditions derived 
from Eq.~(\ref{brane}).  We cannot find a consistent solution 
except for a trivial solution $B_{\mu \nu}$=0,
if one choose the brane action without Born-Infeld term.

\section{Discussion}
\label{sec:discussion}

In this paper we derived the effective theory on the D-brane 
described by Born-Infeld action in type IIB supergravity. To bring out the essence
we focused on the gravity, the U(1) gauge field and the scalars. We considered two 
different derivations by holographic and geometrical approaches. 
In the both, it turns out that the effective theory is four-dimensional 
Einstein+scalar+holographic CFT (or integration of constant) and that the Maxwell fields do 
not appear at the leading order. 
This is a bad news for the D-braneworld scenario. 
Now we have a caution: careful considerations will be demanded in a 
realistic model. 

Usually ``the integration of constant" $\tilde \chi_{\mu\nu}$ in Eq.~(\ref{basic3}) 
is expected to correspond to the holographic 
CFT energy-momentum tensor\cite{GE,GE2}. 
Comparing Eq.~(\ref{basic1}) with 
Eq.~(\ref{basic3}), we can easily confirm that this is the 
case, that is, $\tilde \chi_{\mu\nu}$ is related to $T_{\mu\nu}^{\rm CFT}$ 
at the leading order. More precisely, 
%
\begin{eqnarray}
T_{\mu\nu}^{\rm CFT} & = & \tilde \chi_{\mu\nu}
+\frac{5}{4} h_{\mu\nu} \biggl[ 
\hat D^2 ( -\phi_0 +\rho_0 ) +3(\hat D\phi_0)^2 -\frac{11}{2}\hat D\phi_0 \hat D\rho_0
+\frac{5}{2}(\hat D \rho_0)^2 \biggr] \nonumber \\
& = & \tilde \chi_{\mu \nu} -\frac{1}{4}h_{\mu \nu} J^{(\rho)}_{\rm CFT},
\end{eqnarray}
%
where 
%
\begin{equation}
J^{(\rho)}_{\rm CFT} = \frac{1}{{\sqrt {-h}}}\frac{\delta \Gamma_{CFT}}{\delta \rho}.
\end{equation}
%
In the above we have used the equations for the scalar fields which can be 
obtained through the variational principle of the action 
$S_{\rm D \mbox{-}brane}+S_{\rm ct}+\Gamma_{\rm CFT}$ in the 
previous section
%
\begin{eqnarray}
\hat D^2( \phi_0 -\rho_0 ) -3(\hat D \phi_0)^2-\frac{5}{2}(\hat D \rho_0)^2
+\frac{11}{2}\hat D_\mu \phi_0 \hat D^\mu \rho_0=-\frac{1}{5}J^\rho_{\rm CFT}.
\end{eqnarray}
%
Thus we can confirm the desirable result here. 

As stressed in the Sec.~\ref{IIIB-1}, the current background solution 
is not like AdS spacetime and the gravity cannot be confined on the 
brane at low energy without  compactification. If we compactify the 
extra dimension, we must introduce another brane. In this case, 
the integration of constant $\tilde \chi_{\mu\nu}$ is 
not the holographic CFT energy-momentum tensor but just the 
energy-momentum tensor on the brane. 
If the brane is vacuum, $\tilde \chi_{\mu\nu}=0$. Then the effective 
theory is not like Einstein-Maxwell. But, if the Maxwell field lives on the another 
brane, we can see that the field also appears on the D-brane. 

In this paper we saw the drastic changes from the probe D-brane case when we take
into account 
the self-gravity of the brane. 
Compared to the probe brane, the new ingredients are junction conditions. 
The consistent solutions are extremely limited. In type IIB supergravity, 
indeed, we obtain the consistent bulk solution for the Born-Infeld action, 
but we do not for the Nambu-Goto one. This fact implies that one should be 
careful in connecting an effective action derived from AdS/CFT like correspondence 
to an effective action on a self-gravitating brane. 

There are several remaining studies. The first is 
the higher order corrections and its meaning. In the holographic
approach, we obtained the coupling 
between the curvature and the stress tensor of the gauge fields. 
The systematic analysis will be interesting. 
The second is about the localization of fermion fields 
on the D-brane. We hope that these issues will be addressed in near 
future. 

\section*{Acknowledgments}

We would like to thank Daisuke Ida and Norisuke Sakai for fruitful discussions. 
SO thank Akio Hosoya and Masaru Siino for their continuous encouragements. 
To complete this work, the discussion during and after the YITP workshops YITP-W-01-15 and 
YITP-W-02-19 were useful. The work of TS was supported by Grant-in-Aid for Scientific
Research from Ministry of Education, Science, Sports and Culture of
Japan (No.13135208, No.14740155 and No.14102004). The work of KK was 
supported by Grant-in-Aid for JSPS Fellows.

\end{document}